\documentclass[useAMS,usenatbib]{mn2e}
\usepackage{graphicx}
\usepackage{subfig}
\usepackage{float}

\title[Morphological evolution of the cluster red sequence]{Morphological evolution of cluster red sequence galaxies in the past 9 Gyr}
\author[R. De Propris et al.]
{Roberto De Propris$^{1}$\thanks{E-mail:rodepr@utu.fi} 
Malcolm N. Bremer$^{2}$ and Steven Phillipps$^{2}$\\
$^{1}$ Finnish Centre for Astronomy with ESO, University of Turku, V{\"a}is{\"a}l{\"a}ntie 20, Piikki{\"o}, Finland\\
$^{2}$ H.H.Wills Physics Laboratory,  University of Bristol,  Tyndall Avenue, Bristol, BS8 1TL, United Kingdom}
\begin{document}

\date{}

\pagerange{\pageref{firstpage}--\pageref{lastpage}} \pubyear{2015}

\maketitle

\label{firstpage}

\begin{abstract}

Galaxies arrive on the red sequences of clusters at high redshift ($z>1$) once their star formation 
is quenched and evolve passively thereafter. However, we have previously found that cluster red 
sequence galaxies (CRSGs) undergo significant morphological evolution subsequent to the cessation 
of star formation, at some point in the past 9-10~Gyr. Through a detailed study of a large sample of cluster 
red sequence galaxies spanning $0.2<z<1.4$ we elucidate the details of this evolution. Below $z \sim 
0.5-0.6$ (in the last 5-6 Gyr) there is little or no morphological evolution in the population as a whole, unlike 
in the previous 4-5 Gyrs. Over this earlier time (i) disk-like systems with S{\'e}rsic $n < 2$ progressively 
disappear, as (ii) the range of their axial ratios similarly decreases, removing the most elongated systems 
(those consistent with thin disks seen at an appreciable inclination angle), and (iii) radial colour gradients 
(bluer outwards) decrease in an absolute sense from significant age-related gradients to a residual level 
consistent with the metallicity-induced gradients seen in low redshift cluster members. The distribution of 
their effective radii shows some evidence  of evolution, consistent with growth of {\it at most} a factor  $<1.5$
between $z\sim 1.4$ and $z \sim 0.5$, significantly less than for comparable field galaxies, while the 
distribution of their central ($<1$kpc) bulge surface densities shows no evolution at least at $z<1$. A 
simple model involving the fading and thickening of a disk component after comparatively recent quenching 
(after $z\sim 1.5$) around an otherwise passively evolving older spheroid component is consistent with all 
of these findings.

\end{abstract}


\section{Introduction}

An advantage of studying the cluster galaxy population to test theories of galaxy formation is that, once a galaxy 
has entered a cluster, it will likely remain within a high density environment. We can therefore trace a more direct 
evolutionary connection between high redshift and local systems without needing to appeal to abundance matching 
to link populations at different epochs (e.g., \citealt{Guo10} {\it et seq.}), where the large intrinsic scatter in galaxy growth 
rates in simulations implies that one cannot unambiguously identify galaxy progenitors and descendants between different 
observational epochs in the field \citep{Torrey15}.  The homogeneity of cluster samples offers a snapshot of galaxy properties 
at each epoch, providing a more direct link between galaxy progenitors and descendants. For this reason, cluster galaxies 
may serve as an important benchmark for models of galaxy formation and evolution, as well as being useful cosmological probes. 

The most distinctive and characteristic population  of cluster galaxies at $z<1$ are the cluster red sequence galaxies (CRSGs).
They form such a distinctive population that their colours can be used to search for high redshift clusters in wide-field images
(e.g., \citealt{Gladders00, Eisenhardt08}). These predominantly early type galaxies have little or no ongoing star formation 
subsequent to forming the bulk of their stellar populations and assembling their stellar masses at high redshift (e.g.,
\citealt{Kodama97,Stanford98,DePropris99,Mancone10,Mei12} and references herein). However, while their stellar populations 
and luminosities evolve passively, their morphologies may not, and may also evolve differently from those of similarly passive field 
galaxies.

For example, a population of dense and compact quiescent galaxies has been identified in the field at high redshift and 
appears to be much less common nearer the present epoch (e.g., \citealt{Taylor10, Newman12,McLure13,Damjanov14}). 
Cluster early type galaxies instead appear to have larger sizes than a similarly massive sample of quiescent field galaxies
\citep{Cerulo14,Delaye14} and to evolve more slowly \citep{Kelkar15}. Galaxy size evolution may also have significant 
environmental dependencies \citep{Lani13,Damjanov15}.  In our previous study (\citealt{DePropris15}, hereafter Paper I), 
we explored the morphologies of a sample of CRSGs drawn from four rich clusters at $1<z<1.4$ (mean redshift $z\sim 1.25$). 
We  used deep archival {\it Hubble Space Telescope} imaging  in the rest-frame $B-$ and $R-$bands (observed frame $F850W$ 
and  $F160W$). The latter passband probes the evolution of structural parameters in a rest-frame band more representative of 
the stellar mass distribution (as opposed to earlier studies where the equivalent bandpasses are closer to $U$ or $B$). Our data 
implied even weaker size evolution than previously measured (e.g., \citealt{Cerulo14,Delaye14}) with no significant size growth
compared to the Virgo cluster.

While we confirmed the essentially passive evolution of CRSGs in colour and luminosity, we also observed that a typical CRSG
(with a luminosity around $L^*$ or fainter) in these $z \sim 1.25$ clusters has a considerably more disk-like morphology (lower
S{\'e}rsic index, more flattened light distribution) than its present-day counterparts, structurally closer to that of field spirals. In
agreement with this, \cite{Papovich12} also measured more oblate shapes for CRSGs in a $z=1.62$ cluster partly included 
within the CANDELS dataset, and argued for the presence of increasingly important disk-like components in high redshift 
cluster galaxies. In the field, luminous quiescent galaxies also appear to contain significant disk-like components \citep{
Bundy10,Bruce12, Bruce14, Buitrago13}, with at least some of these systems showing evidence of being rotationally 
supported \citep{Buitrago14}. 

One possible clue to the mechanisms that convert these high redshift `disks' to the more spheroid-dominated population of more
local clusters is offered by the existence of significant radial colour gradients. Local early-type galaxies tend to show slightly bluer 
colours outwards, consistent with the existence of a metal abundance gradient (e.g., \citealt{Saglia00,Tamura00}). Based on the 
measured colour gradients of  \cite{LaBarbera2003} it appears that cluster galaxies are already similar to local early-type galaxies 
in this respect by $z=0.6$. Our sample in Paper I showed significantly larger gradients (also bluer outwards) at $z=1.3$, more
consistent with the existence of age gradients and more similar to the evolution seen by \cite{LaBarbera2003} in their sample of disks. 
Together with the evidence for disk-like components and more flattened shapes (e.g., \citealt{Papovich12}; see Figure 7 in 
Paper I), we argued in Paper I that a population of recently quenched thin disks transforms into the local population of CRSGs 
(dominated by E/S0 galaxies) largely by mechanisms that do not change stellar mass (given the observed passive evolution 
of the luminosity function) or overall sizes (as those of our sample were broadly consistent with local values as measured  
in the Virgo cluster). Possible mechanisms that give rise to these morphological changes include minor mergers, harassment,
fly-bys and tidal effects, among many others.

These objects may also be related to the red spirals found locally in the outskirts of the A901/2 supercluster by \cite{Wolf09} and
in SDSS clusters by \cite{Skibba09} and \cite{Masters10}. Such `anemic' spirals with weak spiral structure and low star formation
rates were originally identified by \cite{vandenbergh76} and \cite{Koopmann98} in the Virgo cluster. These objects may represent 
the low redshift counterparts of our high redshify CRSGs with significant disk components, and would argue that a process of
transformation from field-like spirals to cluster E/S0 may have been taking place in a similar fashion across the past 2/3 of the 
Hubble time. This also suggests that most CRSGs may contain residual disks, a hypothesis supported by earlier photometric 
work of \cite{Michard94} and the recent ATLAS3D dynamical work \citep{Emsellem11,Krajnovic08}.

Here we extend this analysis by bridging the intervening redshift range with the full sample of clusters in the CLASH HST dataset
\citep{Postman12}. Thanks to the multi-band nature of CLASH, we are able to select our targets in the same way as the  targets in 
Paper I , i.e. by rest-frame $R$ band luminosity and $B-R$ colour.  As before, we also measure structural parameters in (restframe) 
$R$ and colour gradients in $B-R$. We can therefore choose objects over similar magnitude ranges and with very similar stellar 
populations at all redshifts (tied by the observed passive evolution in both quantities) and measure them in the same fashion as our 
higher redshift galaxies in Paper I. This yields a comprehensive survey of CRSG structural evolution for objects brighter than 
$\sim M^*+2$  (where M$^*$ is the magnitude corresponding to $L^*$ in the luminosity function), over the last $\sim 9-10$ Gyr, 
drawn from a broad sample of clusters, treated in a uniform manner across redshift (as regards to object selection, use of the 
same passbands and style of analysis).

We discuss the dataset and our analysis in the next section. Results including photometry (in AB system, unless otherwise stated),
measurements of sizes, S\'ersic indices, ellipticities and colour gradients are presented in Section 3.  The results are discussed 
and placed in context {\it via} a simple evolutionary model in Section 4. Conclusions are presented in Section 5. Throughout this 
paper we assume a cosmology based on the latest release of parameters from WMAP9 \citep{Hinshaw13}.

\section{Dataset}

\begin{figure*}
\includegraphics[width=\textwidth]{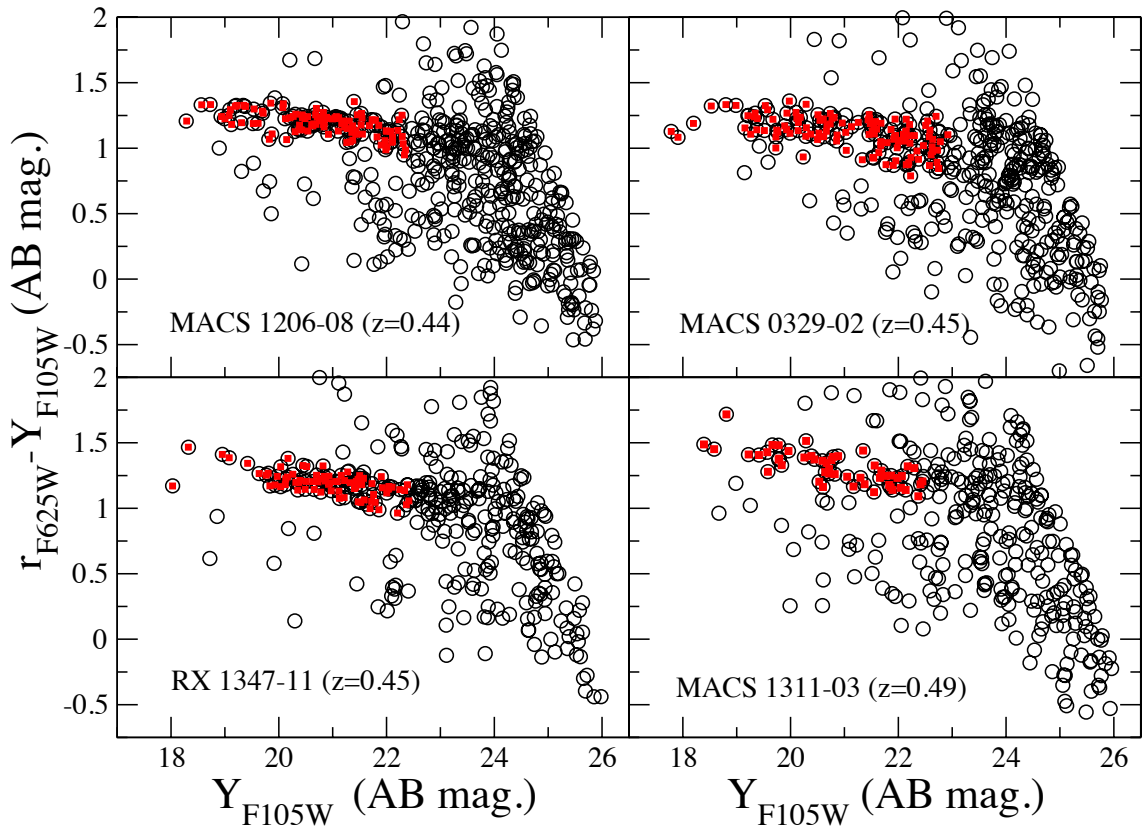}
\caption{An example composite colour-magnitude diagram for the CLASH fields containing clusters in the $0.44 < z < 0.49$
interval, identified in the figure legends. Photometric errors on colour measurements of individual red sequence galaxies  vary 
from $<0.01$ at $Y=19$ to $<0.05$ at $Y=23$. The symbols in this diagram represent all galaxies detected in these fields 
(irrespective of cluster membership). The prominent red sequence in each diagram is formed by passively evolving galaxies 
in each target cluster. Solid red symbols identify the CRSGs analysed to determine their structural parameters. See supplementary
information for other clusters in CLASH and Paper I for the very high redshift sample at $z > 1$.}
\label{examplecmr}
\end{figure*}

The sample of clusters from which all our CRSGs are drawn consists of all available systems targeted in the multi-wavelength 
CLASH Treasury Project\footnote{see \tt https://archive.stsci.edu/prepds/clash} \citep{Postman12}. We supplement this
with four high redshift clusters (XMM1229 -- \citealt{Santos09}, RDCS1252 -- \citealt{Rosati04}, ISCS1434 -- \citealt{Eisenhardt08}
and XMM2235 -- \citealt{Rosati09}) with suitable archival data at $0.98 < z < 1.40$ as already discussed in Paper I. See Table~\ref{tab} 
for basic information. The small number of higher redshift objects in our sample is due to the lack (at present) of suitable infrared
imaging datasets for $z > 0.8$ clusters in the HST archive.

\begin{table}
  \begin{center}
  \begin{tabular}{lll} \hline
    Cluster & Redshift & \\ \hline \hline
    Abell 383 & 0.187 & low \\
    Abell 209 & 0.206 & \\
    Abell 1423 & 0.213  & \\
    Abell 2261 & 0.224 & \\
    RX2129+0005 & 0.234 & \\
    Abell 611 & 0.288 & \\
    MS2137.3-2353 & 0.313 & \\
    RX1532.9+3021 & 0.345 & \\
    RXJ2248-4431 & 0.348 & \\
    MACS1115+01 & 0.352 & \\
    MACSJ1931-26 & 0.352 &\\
    MACSJ1720+35 & 0.391 & \\
    MACSJ0416-24 & 0.396 & \\
    MACSJ0429-02 & 0.399 & \\ \hline
    MACSJ1206-08 & 0.440 & medium \\
    MACSJ0329-02 & 0.450 & \\
    RX1347-1145 & 0.451 & \\
    MACSJ1311-03 & 0.494 & \\
    MACSJ1149+22 & 0.544 & \\
    MACSJ1423+24 & 0.545 & \\
    MACSJ0717+37 & 0.548 & \\
    MACSJ2129-07 & 0.570 & \\
    MACSJ0647+70 & 0.584 & \\ \hline
    MACSJ0744+39 & 0.686 & high \\
    CL1226+3332 & 0.890 & \\ \hline
    XMM1229+0151 & 0.98 & very high  \\
    RDCS1252-2927 & 1.24 & \\
    ISCS1434+3426 & 1.24 & \\
    XMM2235-2557 & 1.40 & \\ \hline
  \end{tabular}
  \end{center}
  \caption{Clusters studied in this paper split into the four redshift dependent
groups discussed in the text.}
  \label{tab}
\end{table}

\subsection{Sample of Clusters}

CLASH clusters were originally selected as representative of the most massive and dynamically relaxed systems at $z > 0.2$ (as
discussed in \citealt{Postman12}), with bright and diffuse X-ray emission, and have typical halo masses in excess of 10$^{14}$ M$_{\odot}$. 
The objects studied in Paper I are also massive clusters, already possessing hot X-ray gas atmospheres, and have richnesses in excess of 
Abell class 1 at $z > 1$; they will evolve into very rare ($\sim 10^{-7}$ Mpc$^{-3}$) systems of richness class above 2 in the present-day 
universe. Unlike previous studies  that select for overdensities in wide-field images (e.g., \citealt{Bundy10,Lani13}), which are 
too small to identify rich systems, our targets are chosen as massive clusters at their respective redshifts, likely among the 
most massive systems in the Universe at their epoch, tracing the highest peaks of the dark matter distribution. For comparison, 
the largest structure studied by \cite{Lani13}, corresponding to the system identified by \cite{Papovich12}, only has a velocity 
dispersion of $\sim 250$ km s$^{-1}$ \citep{Tran15} and does not possess extended X-ray emission \citep{Tanaka13}, indicating a 
far lower environmental density than any cluster in our sample.

\subsection{Photometry and Cluster Membership}

Photometry is carried out exactly as in Paper I, running Sextractor \citep{Bertin96} on the individual images with a set of parameters that 
were determined from previous experience (e.g., \citealt{depropris2013}, Paper I) to be optimal for galaxy detection and photometry. For each
cluster we select the CLASH observations that best match the rest-frame $B$ and $R$ at the respective redshift, as in Paper I, i.e.
the choice of bands depends on redshift. This strategy allows us to directly compare sizes, S{\'e}rsic indices, axial ratios and colour gradients, 
while minimizing uncertainties due to $k$- and $e$- corrections which often affect field studies where structural parameters are determined 
in several different rest-frame bands (cf. the discussion in \citealt{Cassata13} and \citealt{Vika14} on the dependence of structural parameters on
bandpass). All data were retrieved as fully drizzled frames from the Hubble Legacy Archive. All calibrations are on the AB system using the 
appropriate zeropoints from the STScI server. All detections were visually inspected to check for defects, arcs, spurious objects (e.g., fragmented 
bright galaxies) and other contaminants (such as diffraction spikes). Stars were removed using their constant light concentration as a function of 
luminosity. Images in CLASH are already registered to a common frame, but galaxies in the higher redshift sample (Paper I) were position matched 
to within $0.3''$  in the ACS and WFC3 images.

\begin{figure}
\hspace{-0.8cm}
\includegraphics[width=0.55\textwidth]{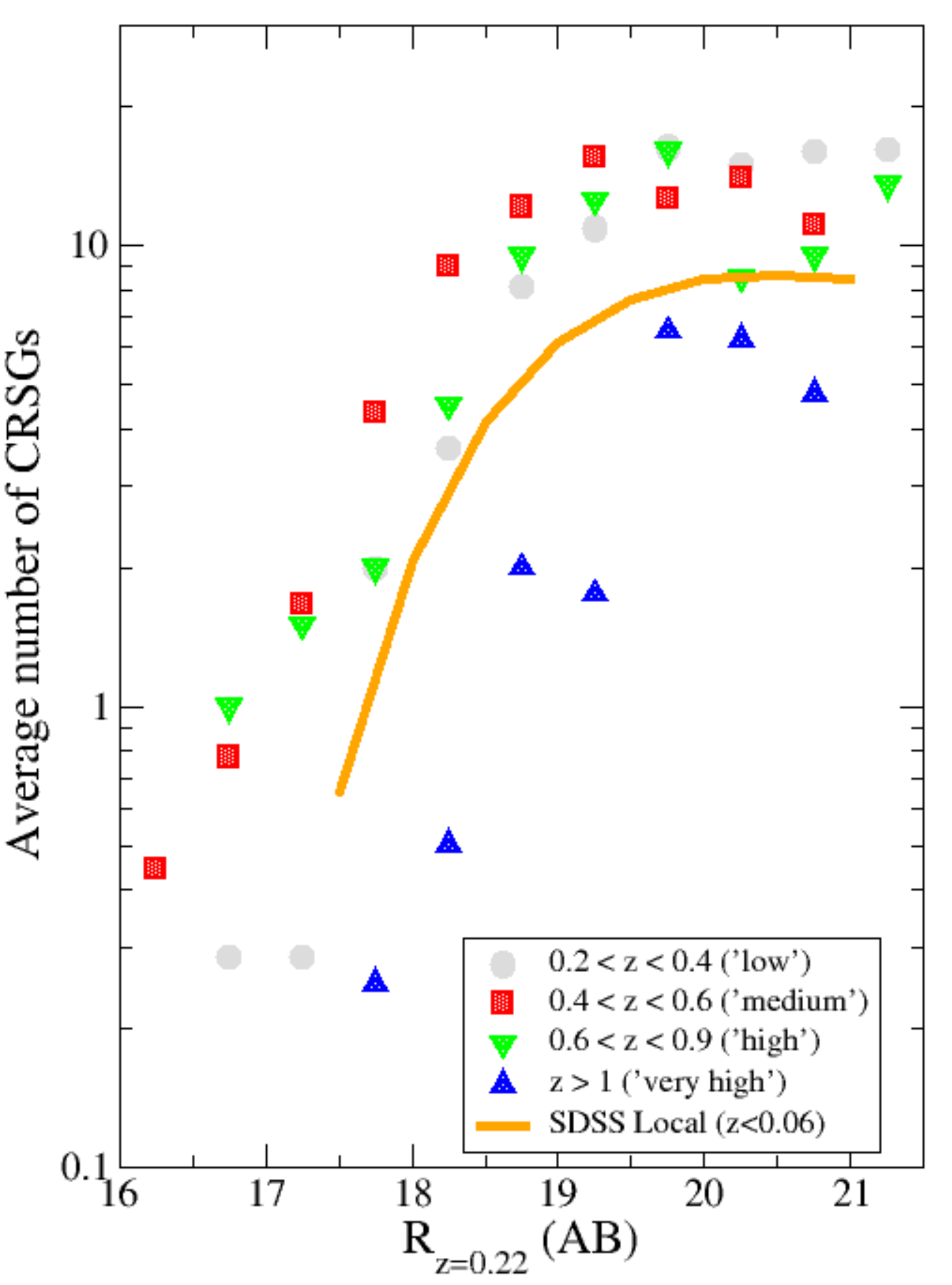}
\caption{The averaged luminosity distributions of galaxies on the red sequences of CLASH clusters (14
clusters for the low redshift sample, 9 for the medium redshift and 2 for the high redshift group), shifted to $z=0.22$ 
(assuming a passively evolving model as described in the text), and our $z > 1$ sample (of four clusters) in 
Paper I (symbols as identified in the legend) compared to the local cluster SDSS $r$ band red sequence
luminosity function as derived by Rudnick et al. (2009), with arbitrary normalization to best show the 
match in shape.}
\label{rsld}
\end{figure}

Within each dataset we observe a prominent red sequence, due to the passively evolving population of massive galaxies
in the clusters targeted, reaching to the photometric limit of the data and showing no apparent deficit of galaxies at low luminosities. 
As with all clusters out to even high redshifts, we observe the expected thin and well-defined red sequences (e.g., the $z=1.8$
cluster by \citealt{Newman14}).These are well separated from their surroundings in colour-magnitude space and therefore serve
to isolate a sample of quiescent cluster members efficiently, especially to the more conservative magnitude limits we adopt for analysis. 
Fig.~\ref{examplecmr} shows the derived colour magnitude diagram in (restframe) $B-R$ vs. $R$ for galaxies in the fields of four 
CLASH clusters at $0.44 < z < 0.49$. For these clusters, the observed  bands are $r_{F625W}$ and $Y_{F105W}$ (at $<z> \sim 0.45$),
equivalent to central wavelengths of 4310 and 7250 \AA\ respectively, compared with 4400 and 6900 \AA\ for the rest-frame $B$ and 
$R$ bands, respectively. Data for the highest redshift clusters are shown in Paper I (e.g., RDCS1252--2927 where the red sequence
is very prominent), while we show the corresponding colour-magnitude diagrams for all other CLASH clusters as supplementary
information (on-line only) in the same fashion as Fig.~\ref{examplecmr}.

We select galaxies within $\pm 0.25$ mag. of the mean colour-magnitude relation for the red sequence, as determined by 
fitting a straight line to the data by linear minimum absolute deviation \citep{Beers90}. In the absence of extensive spectroscopic
information, the  red sequence selects a sample of cluster members with comparatively minimal contamination by galaxies in the
foreground and background (cf. \citealt{Rozo15}). While this means that some contamination may be present, to the limits of our 
analysis its effect is not overly significant. For example,Fig.~\ref{examplecmr}  shows that to $Y \sim 23$ (more than one magnitude 
fainter than our typical analysis limit) the red sequence of a subset of $z\sim 0.45$  clusters appears to be well separated from other 
objects: similar conclusions can be drawn for other clusters (e.g., see the equivalent figure in Paper I). In this and equivalent figures, 
the galaxies we have eventually studied (for structural analysis) are marked as red dots; a few objects in each cluster (even if they 
were on the red sequence) could not be modelled or the residuals seemed to indicate a poor fit. As these represent only a small 
fraction of our sample, we have not considered them further.

The red sequence has been commonly used as an indicator of cluster membership, at least for the dominant quiescent population of
(typically) E/S0 galaxies. As a test of this method, two of our clusters have extensive public redshifts from the CLASH-VLT survey. 
In MACS1206, 62 of our 100 galaxies on the red sequence have a redshift measured from \cite{Biviano13} and Rosati et al. (in preparation). 
Of these 61 are cluster members. In MACS0416, 122 of the 158 galaxies we study on the red sequence also have a redshift from
\cite{Balestra2015} and of these 118 are cluster members. Therefore, our method selects cluster members with a fidelity exceeding 95\%.
Of course higher redshift clusters, and fainter galaxies in any clusters, may suffer from somewhat higher contamination, as we have discussed
in Paper I.  Since the redshift completeness, even for the clusters in CLASH-VLT, is no greater than 60\%, and not all CLASH clusters are
observable from VLT, we still need to use red sequences to select a sample of cluster members in all our clusters.

By construction, our red bands are chosen to match the rest-frame $R$ band, with a small colour term (e.g., as in the example above in 
Figure 2, the central wavelength of the adopted filter is only 5\% different from the actual $R$ band). This colour term is however difficult to
calculate accurately, as it depends sensitively on the spectral energy distributions of galaxies and their evolution. In order to be able to 
compare the luminosity distributions at different redshifts, we corrected all magnitudes to the rest frame $R$ at $z=0.22$ (the mean redshift 
of the five closest CLASH clusters) to provide a reference magnitude, denoted as $R_{z=0.22}$. We assume a passively evolving model 
from \cite{Bruzual03} with $z_{formation}=3$, solar metallicity and an e-folding time of 1 Gyr. Previous work has shown that this model 
yields a good fit to the observed colours of CRSGs across a wide range of redshifts \citep{Mei12}. 

We group the CLASH clusters into three subsamples: (i)  fourteen clusters at $0.19<z<0.40$, the `low' subsample; (ii) 
nine at $0.44<z<0.58$, the `mid' subsample; (iii) two at $z=0.69$ and $z=0.89$, the `high' sample and combine these with the
`very high' $1<z<1.4$ sample from Paper I (see Table~\ref{tab}). The relatively small size of the `high' sample is due to the 
paucity of systems observed in the appropriate bands with HST (ACS F814W  and WFC3 F125W or F160W).  The number of 
CRSGs in each subsample (identified via the red sequence) are 1037 for `low',  743 for `mid', 136 for `high' and 87 galaxies for
`very high' (Paper I) to $R_{z=0.22}=21$. 

\subsection{A $R$-band limited sample of galaxies}

We select galaxies on the red sequence of each cluster and calculate the average number of objects in each of the `low', `medium',
`high' and `very high' redshift subsamples, within 0.5 mag. bins in $R_{z=0.22}$ down to a magnitude where we believe contamination to 
be minimal and red sequences are well isolated in the surrounding colour space (see Fig.~1). 

We plot the average counts (i.e., divided by the number of clusters in each subsample) in Fig.~\ref{rsld} vs. $R_{z=0.22}$, omitting
the error bars to avoid confusion (but the errors are Poissonian, since these are simple counts and do not include the clustering 
contributions from the field). This may be considered as a luminosity {\it distribution} rather than a luminosity {\it function} as we have 
not removed possible contamination from fore/back-ground objects (all clusters are studied in different observed bands, not all of which 
have sufficiently deep, or even any, observations of comparison fields). For comparison, we plot the $z=0.06$ $r-$ band luminosity function 
of cluster red sequence galaxies (corrected to our  bandpass and redshift) from the SDSS \citep{Rudnick09}. The normalisation is chosen 
arbitrarily to best show the differences in the shapes of the red sequence luminosity distribution across the redshift ranges we study and 
to provide a local comparison (this normalisation is usually left to float as it does not have a clear physical meaning for clusters).

The average composite luminosity distributions for the CRSGs drawn from the `low', `mid' and `high' subsamples are consistent 
with the passively evolved Schechter functions for  local and 'very high' redshift clusters having $R^*_{z=0.22} \sim 19.1$ 
(the characteristic magnitude $M^*$ measured in the $R$ band shifted to $z=0.22$ as described above) and $\alpha \sim -1$. 
The mean number of galaxies in the CLASH sample in each magnitude bin are also comparable, suggesting that we are typically
observing objects of comparable environmental density. The number counts of galaxies in the high redshift sample of Paper I are 
also broadly consistent with those measured for CLASH clusters, suggesting that we are not observing objects with very different 
properties across the redshift range of interest.

Given this selection, brighter than an evolved $R_{z=0.22}$ of about 21 (approximately $M_R=-18$, or down to about 1/5th the
luminosity of the Milky Way) we are studying  objects that have broadly the same stellar mass and passively evolving stellar 
populations at all times out to $z \sim 1.25$. The good match to the local luminosity function of \cite{Rudnick09} and to the results of
Paper I confirms the essentially passive evolution of red sequence galaxies in CLASH clusters. While these results are already well 
known, their recovery confirms that our selection of the photometrically-defined red sequence galaxies as cluster members is 
appropriate for the purposes of this study. Therefore, our sample consists of similar galaxies at all redshifts, linked by the observed 
passive evolution on the red sequence. To $R_{z=0.22} \sim 21$ we have a complete sample where we study galaxies of 
similar luminosity and colour at all redshifts with relatively little contamination from non-cluster sources.

\begin{figure*}
\includegraphics[width=0.45\textwidth,angle=0]{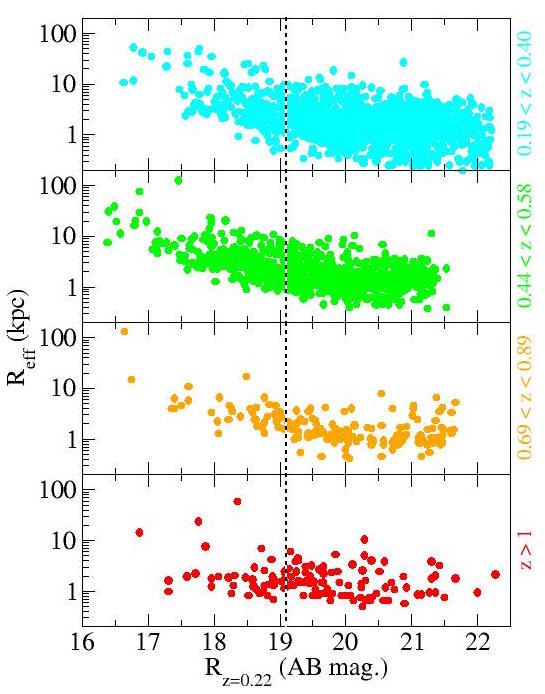} 
\hspace{1cm}
\includegraphics[width=0.45\textwidth,angle=0]{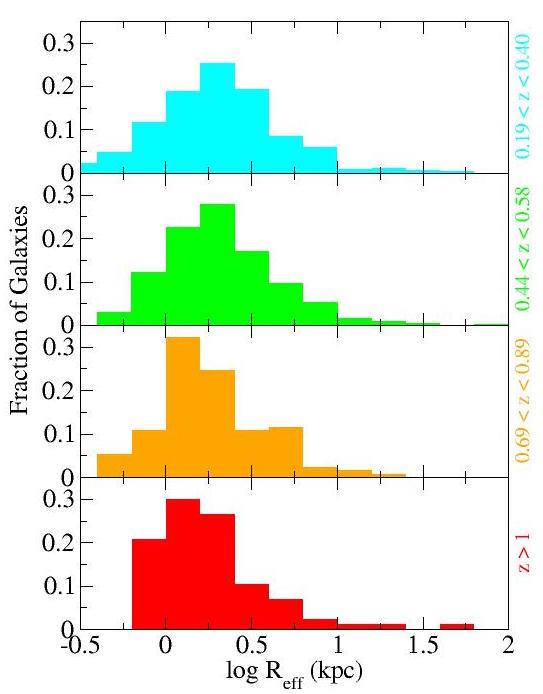} 
\caption{{\it Left:} Size distribution for galaxies in clusters. We plot the effective radii from single S\'ersic
profiles vs. the $R$ magnitude at $z=0.22$ (where this is actually the HST band closest to
rest-frame $R$, accounting for passive evolution and cosmology) for groups of clusters at `low'
($0.19 < z <0.40$), `mid' ($0.44 < z < 0.58$) and `high' ($0.69 < z < 0.89$) redshifts in the CLASH
sample. We also plot the $z > 1$ (`very high') results from Paper I in the bottom panel (as indicated in the
figure). The errors returned by GALFIT on these parameters are typically below 1\% and smaller than
the size of the points. The gray line shows the $M^*$ value derived by Rudnick et al. (2009) for
local clusters in the $R$ band, corrected to $z=0.22$ with the concordance cosmology and the passive
evolution model (i.e., $R^*_{z=0.22}$)  {\it Right:} Size distribution for galaxies in clusters at all redshifts 
we consider. Each panel corresponds to a redshift interval as indicated in the legend. The bottom panel 
shows the data from Paper I for the `very high' ($z > 1$) sample}
\label{sizes}
\end{figure*}

\section{Structural parameter evolution}

We determine structural parameters for the CRSGs in the rest-frame $B$ and $R$ using {\tt GALFIT} \citep{Peng02,Peng10} as 
in our previous analysis of $z \sim 1.25$ clusters in Paper I. 

For each galaxy we carry out an interactive and iterative process, where we adopt reasonable starting values and proceed to
inspect the model and residuals to ascertain the quality of the fit (i.e., we do not run GALFIT in batch mode, but individually for
each object). Where necessary, multiple models are run together, e.g., in the case of close pairs or systems of galaxies. We adopt  
a single S\'ersic profile. This is warranted by the relatively shallow CLASH data and can also be more easily compared to previous 
work in cluster and field galaxies. For each image and in each band we use the median of several bright, unsaturated and isolated 
(as possible) stars to derive a PSF (point spread function) reference, with min/max clipping of eventual interlopers in the chosen field. 
Typical PSF image sizes are 100 times the FWHM of the PSF (as recommended by the GALFIT manuals). This procedures ensures that 
the PSF incorporates all processing steps to which the images of the galaxies  have been subjected and is therefore preferable to 
generating artificial PSFs (e.g., with TinyTim). For each galaxy being considered the sky level is determined locally within the large 
aperture needed for analysis by GALFIT (including assessment of eventual gradients by fitting a surface to the sky -- see below). 
The results of this procedure are the effective radii (in restframe $B$ and $R$), the S\'ersic indices (also in both bands) and the
axial ratio b/a (also in $B$ and $R$ restframe). From these we derive the colour gradient $\Delta(B-R)/\Delta \log r$, following the 
formula given in \cite{LaBarbera2002,LaBarbera2003} using the measured structural parameters in each band.

Particularly for the higher redshift samples, the pixel scale of the images ($0.03"$ to $0.08"$) is a significant fraction of the effective
radius (at $z=1$, $1"$ corresponds to 8 kpc, with typical galaxies having $R_{eff}$ of 1 kpc, e.g., \citealt{Gutierrez04}) and the PSF
Full Width at Half Maximum (FWHM) has a size comparable to the effective radius ($0.05"$ to $0.08"$ depending on the bandpass
and instrument used). Any 2D profile, e.g. produced with the IRAF {\tt ellipse} package \citep{Jedrzejewski87}, will be heavily affected
by sampling and by the PSF size. Furthermore, mismatches in the galaxy centroids and the different PSFs in the red and blue bands
(often using two different instruments: ACS in the optical and WFC3 in the infrared) will result in spurious colour gradients (e.g.,
\citealt{LaBarbera2003}). Using fits to structural parameters from GALFIT modelling of the full image, automatically removes these
sources of error. We therefore use our derived parameters in the subsequent analysis rather than attempting more detailed 
profiling.

It is known that the errors output by GALFIT can sometimes be significant underestimates in the case of small images. 
However from the work of \cite{Haussler2007} who simulated HST images very similar to those used here, in terms of magnitude, 
surface brightness and size, it can be seen that GALFIT will provide values of $R_{eff}$ accurate to within 5\% down to magnitude 22.5, 
even for steep ($n \sim 4$) profiles. Errors in $n$ are $\sim 0.25$ at $n \sim 4$ and much less at $n \sim 1$, while ellipticity errors are
 typically 0.04. Errors of this size, for the faintest objects we analyse, are sufficiently small that they do not influence any of our results.

In addition, we have run a series of simple simulations of our own with IRAF {\tt artdata} to test our procedures. These introduce idealised galaxies 
(pure exponentials or pure spheroids obeying the deVaucouleurs law) in `clean' regions of the images. Using GALFIT on the simulated images 
confirmed that parameter errors were of the level noted above or smaller (the dispersion in values was consistent with the errors returned
by GALFIT, i.e., within 1\%) and that GALFIT parameters could be used safely down to the faint limit of our analysed data. 

We have further introduced artificial sky gradients to simulate the effects of intracluster light and GALFIT is found to be insensitive to 
any such smooth gradients, unless they vary rapidly across the galaxies over scales comparable to the effective radius; even then, 
the result is not incorrect parameters but catastrophic failures of the iteration. When this does not happen, the errors 
returned are consistent with the GALFIT output, i.e., $\sim 1-2\%$. Since we actually inspect each object (model and residuals) 
individually, we can rule out the presence of such pathological sky features at any levels that would interfere with the fitting procedure.

\section{Results}

\subsection{Size evolution}

Fig.~\ref{sizes} shows the size distribution, vs. $R$ band magnitude (shifted to $z=0.22$ assuming the concordance cosmology
and a passively evolving model as described above), for all galaxies in the CLASH cluster sample and the sample from Paper I. 
The appearance of the distribution is similar to that found in local systems (e.g., Coma in \citealt{Gutierrez04}), with a cloud of
systems at intermediate sizes of $1$--$2$ kpc and a brighter sequence of large, luminous galaxies; these latter objects tend to be 
high $n$ early-type spheroid-dominated galaxies (classical E/S0). We compare the size distributions of galaxies brighter than
$R_{z=0.22}=21$ in the right-hand panel of Fig.~\ref{sizes} (including the $z > 1$ sample in Paper I).

In order to take into account both the shape and absolute numbers of objects contributing to each distribution in 
Fig.~\ref{sizes}, each was compared to the others using a standard Kolmogorov-Smirnov (KS) test. The `very high' 
distribution differed from the `low' and `middle' at the 3$\sigma$ and 2$\sigma$ levels ($P=0.02$ and 0.06) respectively (assuming 
a Gaussian distribution) i.e. unlikely to be drawn from the same distribution. Similarly, the `high" distribution differed from the `low' 
and `mid' distributions at the $\sim 3\sigma$ (P=0.02 and 0.04) levels respectively. Below $z \sim 0.6$ there appears to be no 
significant size evolution, while for higher redshifts,  we can characterise the change by the typical $R_{eff}$  decreasing  by {\it at 
most} a factor $\sim 1.5$ over $0.7 < z < 1.4$ given that the peak of the size histogram shifts by at most one bin (0.2 dex) in
Fig.~\ref{sizes}. The large sizes of some of our brightest cluster galaxies are also consistent with the findings of \cite{Stott10} and 
\cite{Lidman11}. \cite{Stott10} use both a method very similar to ours (with GALFIT) and also the IRAF {\tt ellipse} package from 
{\tt sdsdas}, finding very similar results, suggesting that the measured sizes are insensitive to the precise modelling technique
 (J. Stott, private communication)

The results shown in Fig.~\ref{sizes} are broadly consistent with the form of size evolution identified by \cite{Delaye14} from
ground-based near-IR imaging, through a comparison between systems at $0.8<z<1.5$ and $z=0$ ($R_{eff}\propto (1+z)^{-0.5}$)
with most, if not all of the evolution, taking place well above $z \sim 0.6$ and probably beyond $z \sim 1$.  Any size evolution in 
CRSGs in the past $\sim 9$ Gyr appears to be much smaller than that found in field samples of quiescent galaxies of comparable
luminosity over the same redshift range (\citealt{Saracco14,Cerulo14, Delaye14}, Paper I). \cite{Kelkar15} observes that at $0.4
< z < 0.8$ early-type galaxies have roughly the same size as local ones, in agreement with our findings for the `low', `mid' and `high'
subsamples. \cite{Newman14} detects no more than a 30\% change in the size of red sequence galaxies with stellar masses
higher than $10^{9.8}$ $M_{\odot}$ between $z=0$ and $z=1.8$ (also see a forthcoming paper by S. Andreon, private communication).

\cite{vanderwel14} and \cite{Huertas15}, among others, however find that the average field early type galaxy increases its 
$R_{eff}$ by a factor of 3 or more, mainly since $z\sim 1.5$, with growth continuing all the way to $z=0$, unlike this cluster 
sample. The degree of size evolution appears to depend on environment: \cite{Lani13} and \cite{Damjanov15} 
also find  evidence that size evolution is weaker in high density environments (by as much as 50\% in \citealt{Lani13}
when looking at their densest systems). However, note that our clusters are considerably more massive than any of 
the objects studied by \cite{Lani13} or \cite{Damjanov15}. Given the already known passive evolution in colour 
and luminosity, the small change in size observed for these galaxies appears to suggest early formation and assembly, with 
little subsequent evolution, especially by mergers, and most other changes being due to internal processes or more gentle 
external inputs, a conclusion also reached by \cite{Huertas15} at least for the more massive spheroidals in the field. 

\subsection{S\'ersic index evolution}

\begin{figure*}
\includegraphics[width=0.45\textwidth,angle=0]{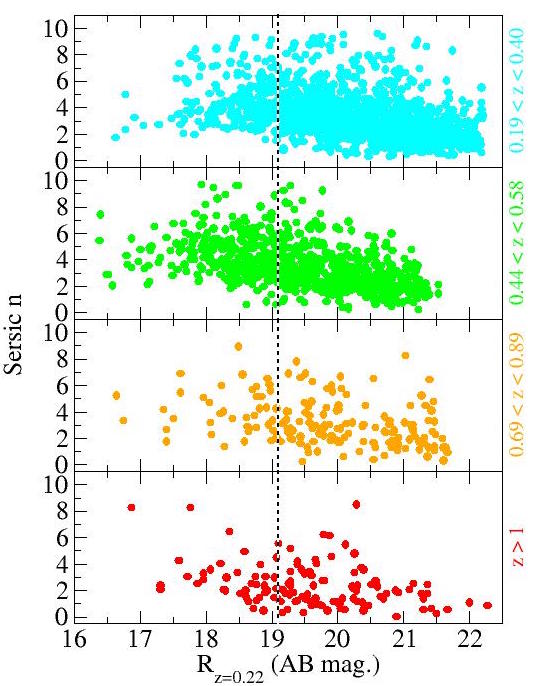} 
\hspace{1cm}
\includegraphics[width=0.46\textwidth,angle=0]{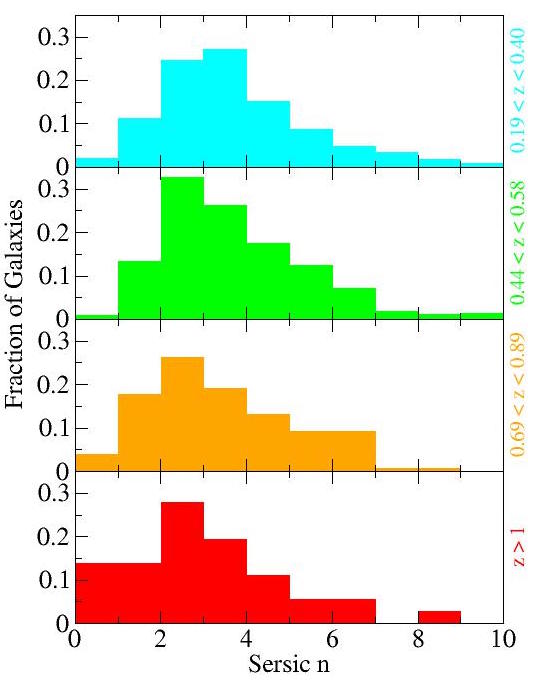} 
\caption{{\it Left:} Distribution of S\'ersic indices for galaxies in clusters. We plot the S\'ersic index $n$  vs. the $R$ 
magnitude at $z=0.22$ as in the previous Figure 3, for groups of clusters at `low' ($0.19 < z <0.40$), `mid' ($0.44 < z < 0.58$)
and `high' ($0.69 < z < 0.89$) redshift in the CLASH sample. We also plot the $z > 1$ results from Paper I (`very high') in the 
bottom panel. As in the previous figure, the thick dashed line shows the predicted position of the $R^*_{z=0.22}$ point based on the 
luminosity function of Christlein \& Zabludoff (2003). Errors returned by GALFIT are smaller than the size of the points (typically 
$< 0.01$ in $n$). {\it Right:} Sersic index distribution for galaxies in clusters as in the previous Figure 3 and indicated
in the legend. There is evidence for an increasing presence of disk-like components among red
sequence cluster galaxies as a function of lookback time.}
\label{Sersic}
\end{figure*}

As in the previous subsection, we now plot the evolution of rest-frame $R-$band S\'ersic indices ($n$) as a function of redshift for 
CRSGs in Fig.~\ref{Sersic}. The corresponding histograms for the distribution of $n$ values is shown in the right-hand panel of
Fig.~\ref{Sersic}, including the  $1<z<1.4$ data in Paper I.

At low redshift ($z<0.6$), clusters are dominated by a population of spheroidal galaxies, with $n > 2.5$ corresponding
to classical E/S0 galaxies as expected, with very few objects with $n<2$.  However, at higher redshifts an increasing fraction 
of the population have low S\'ersic indices consistent with  the presence of disk-like components (Fig.~\ref{Sersic}). Whereas 
for $z < 0.6$ about 30\% of galaxies are observed to have $n < 2.5$ and the median $n$ is $\sim 3.3$, this fraction rises 
slightly (to about 35\%) in the $0.6 < z < 0.9$ sample and in the highest redshift subsample more than 50\% of galaxies have 
$n < 2.5$ and the median $n$ is 2.2, with only brighter cluster members tending to be $n > 4$ spheroids. 

Again, KS-tests were carried out between each of the subsamples. The S{\'e}rsic index distribution of the `very high' subsample 
differed from each of those of the other three subsamples at more than the 5$\sigma$ level (P$<$0.01). The `high' and `mid'  
distributions showed no convincing differences from each other, but both differed from that of the `low' subsample at the approximately
2$\sigma$ (P=0.04) level, implying ongoing evolution towards the low redshift distribution to about $z \sim 0.5$ -- $0.6$. Looking at
Fig~\ref{Sersic} any evolution below $z \sim 1$ is likely to be more subtle than that at higher redshifts and may only require 
evolution in a subset of the galaxies to achieve the observed differences. 

\begin{figure*}
\includegraphics[width=\textwidth,angle=0]{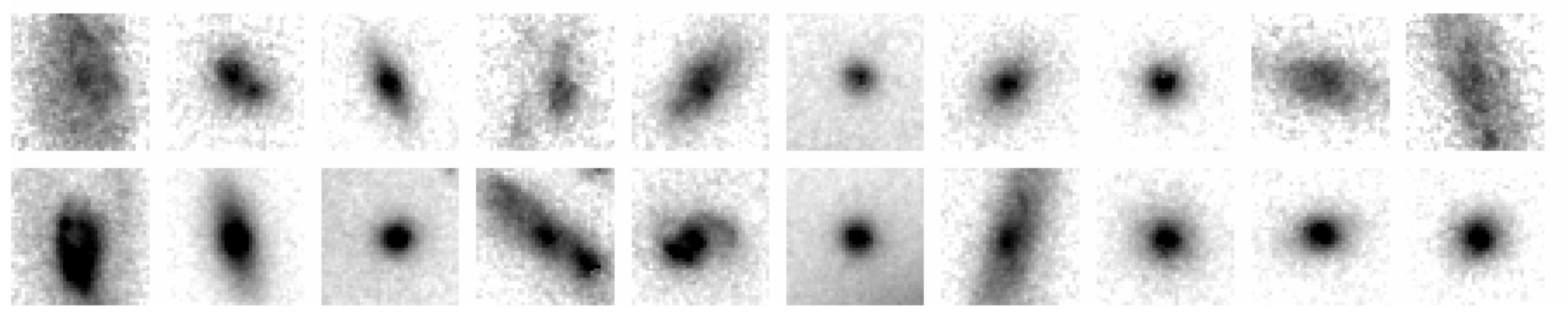}
\caption{HST/ACS $2 \times 2"$ $I$-band images for 20
red sequence galaxies with $n<2$ and magnitudes within 
$\sim 1$ magnitude of $M^*_R$ from the $z=0.9$ cluster CL1226+3332. 
Presentation is the same as that for figure 7 in Paper 1 with the  
grey-scale stretching between the mean sky level to the square-root of 
the central surface brightness (so more diffuse, lower surface brightness 
galaxies appear to have noisier images).}
\label{cl1226}
\end{figure*}

Inspection of the $B$ band images confirms this impression. In Paper I (see Fig.~7) we saw that most $n < 2$ galaxies have 
significant disk-like components, even though they lie on the red sequence. In this paper, we plot similar postage stamps of 
all such galaxies in CL1226+3332 ($z=0.89$), the highest redshift cluster in the CLASH sample, in Fig.~\ref{cl1226}, showing 
that disk-like objects start to become significant at this redshift, including some slightly clumpy systems. Both \cite{Rembold12} 
and \cite{Cerulo14} also observe numerous faint disk-like galaxies in two $z=1$ clusters.

This reinforces our previous finding: red sequence galaxies in high redshift clusters appear to contain an increasingly important 
disk-like component at higher lookback times. While local S0 galaxies certainly have disks, their overall S\'ersic indices are still 
typically $n\ge2.5$. Consequently this disk-like component at high redshift must eventually either disappear or evolve by $z \sim 0.6$ 
where the population closely resembles that of present-day clusters. The end-state of these high redshift CRSGs (with significant
disks) must be the S0 and elliptical galaxies that dominate at higher $n$ and lower redshifts. In some respects, this finding resembles
the observations by \cite{Buitrago13} in the general field, where apparently quiescent galaxies exhibited an increasing contribution 
from disk-like systems at $z >1$. Similarly, \cite{Huertas15} observe bright $n\ge4$ spheroids formed at high redshift, and an 
increasing proportion of irregular and thin disks at lower luminosities and higher redshifts. In order to ascertain the nature of the 
disk-like components we next look at the shape of the objects via their axial ratios.

Below $z \sim 0.6$ our clusters contain mainly, if not solely, galaxies with effective radii, S\'ersic indices and axial ratios (see below)
consistent with $z=0$ systems. Such objects likely resemble conventional E/S0 galaxies (to the extent that such classification is 
objective and appropriate, considering the nature of the data). We do not carry out a study of such classifications here, as we are
interested in the change in structural parameters and quantifiable measures of evolution.

\subsection{Axial ratio evolution}

\begin{figure*}
\includegraphics[width=0.45\textwidth,angle=0]{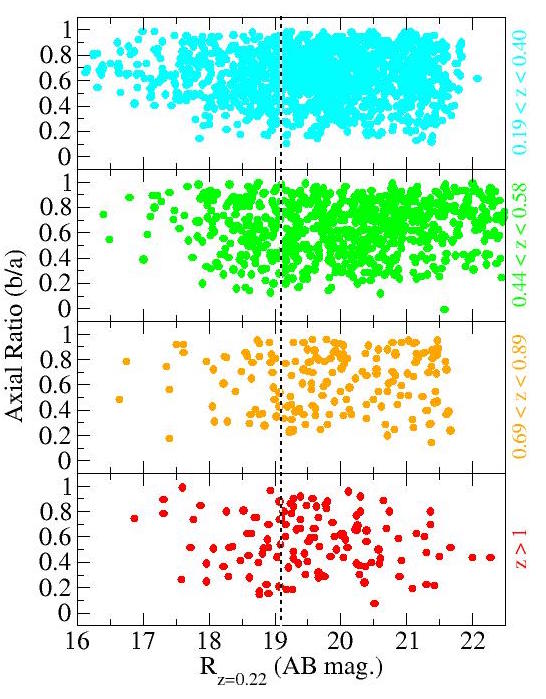} 
\hspace{1cm}
\includegraphics[width=0.462\textwidth,angle=0]{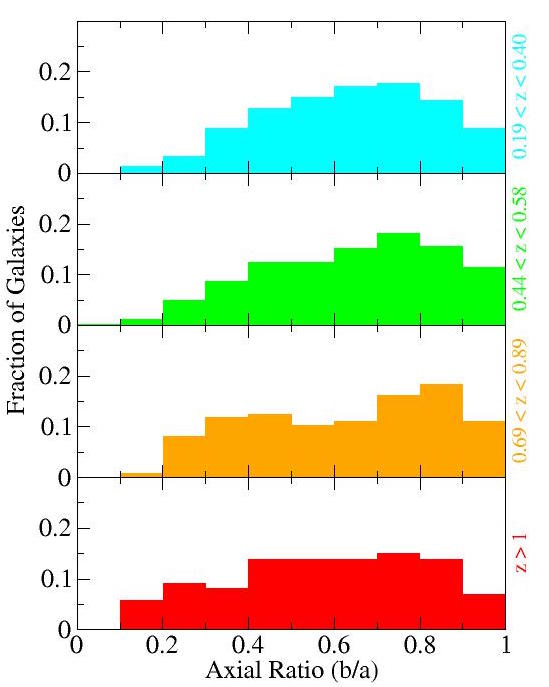} 
\caption{Distribution of axial ratios for galaxies in clusters in the previously defined subsamples. Errors on axial ratios returned 
by GALFIT are smaller than the size of the points {\it Right:} Axial ratio distribution for galaxies in clusters as in Figure 4 (see 
legend). There is evidence for an increasing disk fraction (oblateness) among a subset  of the CRSGs as a function of lookback 
time.}
\label{axial}
\end{figure*}

In Fig.~\ref{axial} we plot the distribution of b/a (axial ratios) vs. $R_{z=0.22}$ magnitude for all our 
clusters in the same fashion as for sizes (Fig.~\ref{sizes}) and S\'ersic indices (Fig~\ref{Sersic}). We also 
plot the corresponding axial ratio distributions in Fig.~\ref{axial}.

At $z<0.6$  the axial ratio distributions indicate a population dominated by objects with low apparent ellipticity, either true ellipticals 
or systems with comparatively thick disks such as S0s. The distribution peaks close to the local ($z=0$) values in \cite{Ryden01},
i.e. $b/a \sim 0.7$--$0.8$. The axial ratio distributions of the two higher redshift populations instead appear by eye to be flatter, 
with a significantly higher proportion of systems with projected axial ratios reaching down to $\sim 0.2$. Such distributions are 
unlike those achievable through a mixture of ellipticals and typical S0 galaxies \citep{Lambas1992} and again indicate an 
additional population with significant disk-like components at $z>0.6$ with an axial ratio distribution closer to that of field spirals
\citep{Graham08}: the mean axial ratio in the highest redshift bin is 0.58, closer to that observed in spiral disks \citep{Ryden04}. 
KS-tests indicate significant differences between the `very high'  and both the `mid'  and `low' subsamples at about the 2$\sigma$ 
levels (P=0.05 and 0.09 respectively), but no other significant difference. 

\cite{Buitrago13} also finds some evidence for increasing flattening with redshift, implying a more prominent role for disk-like
components in galaxies, while \cite{Chang13} find an increase in the fraction of oblate objects at $z > 1$. However, unlike their 
results we have an increasing fraction of flatter galaxies at the lower luminosity end in the higher redshift bins. \cite{Papovich12} 
already noted the presence of  elongated galaxies on the red sequence for a $z=1.62$ cluster, while \cite{Holden09} has found 
no change in the shape of cluster galaxies {\it selected as S0} with redshift, which implies that our flattened `disks'  might 
eventually thicken and become rounder, in order to join the population of present-day S0s.

\subsection{Colour gradients evolution}

A further clue to the evolution of galaxies may be provided by radial colour gradients. These reflect the 2D  distribution of stellar
populations as well as their kinematics. The existence of colour gradients in nearby ellipticals (e.g., \citealt{Saglia00,Tamura00})
is best explained as being due to a mild metal abundance gradient with radius (richer inward) and its apparent dependence on 
galaxy luminosity (e.g., in Coma: \citealt{DenBrok11}) suggests a model where the majority of the stellar populations were 
assembled early in rapid mergers or dissipational collapses, as a more extended hierarchical merger scenario would tend to 
erase such gradients \citep{Kobayashi04}. 

In Paper I we showed through the ratio of rest-frame $B-$ and $R-$band $R_{eff}$ that the $1<z<1.4$ CRSGs had significant radial 
colour gradients, becoming bluer with radius. The strength of these gradients was such that they could not be explained by abundance
variation, implying that they were a symptom of radial age gradients in the stellar population of each galaxy. Following Paper I, we have
derived colour gradients for our galaxies, extending the treatment by converting the observed $R_{eff}$ and $n$ in the two $B$ and $R$
bands to a more conventional $\Delta (B-R) / \Delta \log r$ \citep{LaBarbera2002}. To do this we determine the colours at 1 and
2$R_{eff}$, sufficiently large radii to not have the measurement compromised by either the PSF or pixel size, even for the most 
distant objects. 

\begin{figure}
\hspace{-0.6cm}
\includegraphics[width=0.55\textwidth]{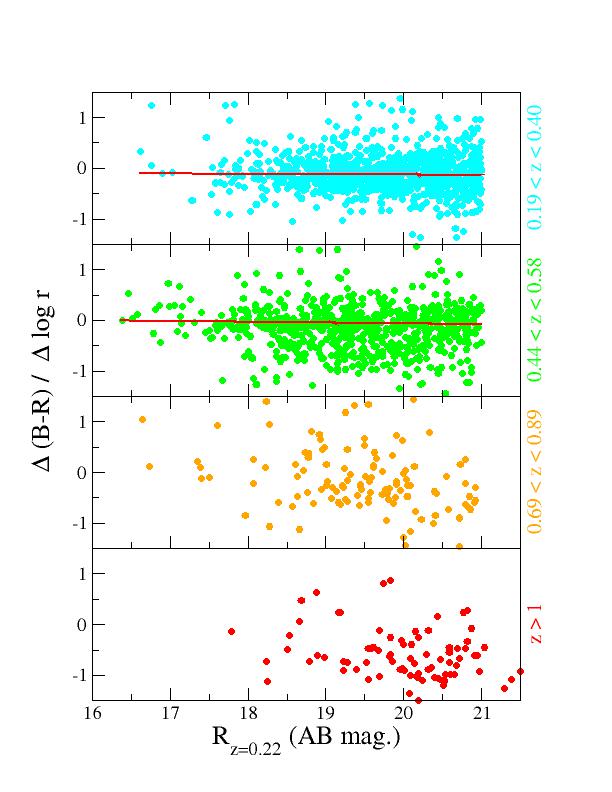}
\caption{Colour gradients for galaxies in clusters at all redshifts we consider. Each panel corresponds to a redshift interval as 
indicated in the legend.}
\label{gradients}
\end{figure}

In order to display the results of this procedure and gain insight into how the gradients depend on stellar mass, for each redshift range 
we show the derived colour gradients and best fitting straight line from a biweight estimator (e.g., \citealt{Beers90}) in
Fig.~\ref{gradients}, plotted vs. $R_{z=0.22}$ as in previous figures. The units of the colour gradients are magnitudes per decade 
in radius. The statistical uncertainty on each derived value is low (typically $<0.1$ mag per decade in radius) given the errors on 
$n$ and $R_{eff}$ from GALFIT (typically less than 1\%). Consequently the scatter in and breadth of the distributions is a real reflection 
of the scatter in the properties of the objects at each redshift.

At lower redshifts, the mean values of the colour gradient at each $R-$band magnitude are consistent with the usual mild colour
gradients (about 0.1 -- 0.2 mag. blueward per decade in radius) seen in nearby clusters. At these redshifts we also confirm the 
existence of a mild trend with luminosity (and presumably stellar mass) extending over 3--4 magnitudes, in the sense that 
fainter galaxies have milder colour gradients, as seen in the Coma cluster \citep{DenBrok11}. This relationship appears to persist 
at least out to $z \sim 0.6$ in agreement with earlier studies by \cite{LaBarbera2003}. Linear fits (with minimum absolute deviation) 
to the data are shown for the `low' and `mid'  redshift bins in Fig.~\ref{gradients}. If we interpret the observed gradients as metal
abundance gradients, then our data are consistent with a trend of metal abundance with stellar mass, in the sense of galaxies
with steeper potential wells having more significant gradients, as in Coma \citep{DenBrok11}. This is more consistent with an inner 
region dominated by old stellar populations formed dissipationally at high redshift, either by gas-rich early mergers or large-scale
gaseous collapses (e.g., \citealt{Oser10}).

The picture changes at higher redshifts. At $z\sim 0.2$ the gradient is typically -0.1 mag  at $R_{z=0.22}=20$ while at $z>0.6$
this becomes  --0.5 mag, but with a larger scatter. At $z>0.6$ the outskirts of galaxies seem to become increasingly blue relative 
to their centres and show an increasing object-to-object scatter in the strength of the gradient (cf. Paper I,\citealt{Chan2016} 
for a similar result). For this reason we do not carry out fits to the data in the `high' and `very high' bins in Fig.~\ref{gradients}. 
The appearance of comparatively strong colour gradients at $z>0.6$  is  consistent with the results at $z\sim 1.25$ in Paper I 
where the outskirts of the CRSGs were clearly bluer than their centres (even though they  were globally red in colour by selection). 
At these redshifts a significant minority of sources with $R_{z=0.22}>19$ appear to have gradients with  strength more easily achievable 
through stellar age gradients, rather than variations in metallicity as is common at lower redshifts. 

In order to explore how the variation in the strengths of  colour gradients maps onto potential differences in both age and metallicity 
of stellar populations in bulge and disk components, we generated the luminosity and colour evolution with redshift for two possible
components of the galaxies using \cite{Bruzual03} SSP models. We assumed a component that  formed in a single burst at $z=3$ 
with Solar metallicity and determined its variation in rest-frame $B-R$ colour and mass to light ratio at $z<1.5$. This can be taken 
to be representative of a maximally old bulge component (or indeed a similarly old disk component).  For purposes of this exercise 
we assume that this represents  a maximally red model for the central regions of one of our galaxies. We then did the same for 
a component undergoing a similar burst at $z=1.5$ with either Solar or 0.25 Solar metallicity. The lateness of this burst was 
chosen simply to give the maximally blue colour and most rapid colour evolution below $z=1.4$ rather than attempting to mimic 
any individual galaxy. In effect, it places an upper  bound on the variation in colour  that can be attributed to variations in population 
age and metallicity across a galaxy. 

\begin{figure}
\hspace{-1cm}
\includegraphics[width=0.45\textwidth, angle=-90]{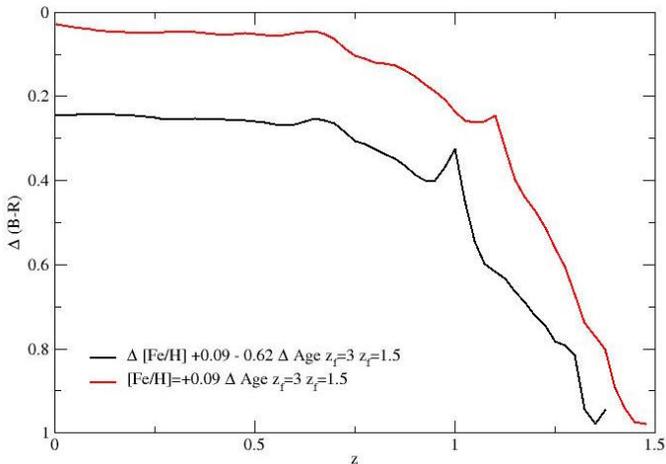}
\caption{Difference in colour between a Solar metallicity SSP burst at $z=3$ and two SSP bursts at $z=1.5$ when viewed at $z<1.4$. 
The two younger SSPs have metallicities of approximately 0.25 and 1 times Solar. These models illustrate the most extreme 
colour differences likely to occur in a quenched system with two different stellar populations.  The oldest burst represents a 
bulge population and the youngest a disk. In reality both are likely to consist of stars from more extended periods of star 
formation, decreasing colour differences between them relative to those derived here.  Nevertheless, the models demonstrate 
that the colour gradients observed in the CRSGs studied here  can plausibly be explained by a mix of age and metallicity-related
gradients, with the age-induced gradients dominating at earlier times over the metallicity gradients.}
\label{models}
\end{figure}

The range of colour differences between the old and younger stellar populations, and its evolution with redshift demonstrates that 
the variation and evolution in colour gradients observed in the population of  CRSGs can be potentially explained through a 
combination of metallicity and age-related gradients. The age-induced gradients decrease with time as both populations 
converge towards similar colours $\sim 5$ Gyr after the most recent burst, revealing the underlying gradient due to the (known)
radial metallicity gradients (see also \citealt{Chan2016} for a similar conclusion). While similar colours may in principle be achievable 
if the younger component included some ongoing star formation below $z=1.4$, this behaviour is constrained by the lack of evidence 
for such activity in cluster red sequences below this redshift. The differential colours shown in Fig.~\ref{models} are for two components 
of equal stellar mass. Given that specific star formation rates in these galaxies are usually well below $10^{-10}$\ yr$^{-1}$, it is difficult 
to see how the colour gradients observed at $z>0.6$ are achievable through ongoing star formation alone. 

A similar argument is presented by \cite{Chan2016} who analyse one of the clusters we studied in Paper I and reach consistent 
conclusions to ours: the presence of an age gradient superposed over a more ancient weak metallicity gradient, although they do
not make the explicit connection to the presence of disk-like components. Studying red disks in the SDSS \cite{Lopes16} also 
argue for a quenching timescale of around 2--3 Gyr (in agreement with our work and \citealt{Chan2016}) and the gradual morphological
evolution of the red disks as they move towards cluster centres.

\section{Discussion}

Paper I explored the morphological differences between a sample of $z>1$ CRSGs and their nearby counterparts. A significant 
fraction of  the reasonably bright  ($\sim L^*$) high redshift systems had  structural parameters that were consistent with more 
disk-like morphologies than those of their low redshift counterparts. The brightest galaxies in the red sequence appeared to be 
different, in that their morphological parameters were already closer to local early type ellipticals, indicating any structural 
evolution was completed at earlier epochs for this brighter subsample. What was not established in Paper I was the timescale 
over which any morphological  transformation occurred in the more typical red sequence galaxies. The results within this paper 
help to determine this timescale by exploring the same structural parameters for CRSGs drawn from clusters observed at several
intervening epochs.   
 
A key result of this work appears to be that for most of the structural  parameters in question, any strong evolution has largely 
been completed by $z\sim 0.5-0.6$, a look-back time of 5--6 Gyr, and 3--4 Gyr after the latest redshift ($z\sim 1.4$)  that any 
star formation in the red sequence galaxies is likely to have been quenched \citep[e.g.][]{ Tran2010}. At higher redshifts there is 
good evidence (albeit limited by the number of systems with useful HST data at $0.6<z<1$) that the relevant  parameters are
transitioning between the local distributions and those found in Paper I. As the velocity dispersions are already very high in these
massive clusters and the luminosity function evolution disfavours significant major merging, this process must take place 
secularly or by more gentle external inputs (such as minor mergers or fly-bys).

All of this data is consistent with a simple picture  outlined in Paper I where, after quenching, $\sim L^*$ red sequence
galaxies do not  evolve in colour and luminosity other than the passive evolution of their stellar populations, but a proportion 
undergo considerable morphological change over the past 2/3 of the Hubble time. The brightest and more massive cluster 
spheroids in this sample seem to have already arrived at their final configuration by the beginning of this period (cf., \citealt{
Saracco14}) but fainter galaxies are likely to show significant evolution, analogous to the observations by \cite{Huertas15} 
in the CANDELS field. However, most of our discussion as to the former must be qualitative for now, as there are too few 
bright galaxies in the small number of high redshift clusters that we observe to draw statistically valid conclusions.

Within this picture, the CRSG population as a whole appears to become slightly less compact with time, although the scale of 
this growth is significantly less than that seen for passively evolving field galaxies of similar mass over the same time period.  
The evolution is reminiscent of that observed in previous studies \citep{Delaye14}, although it may be even weaker, compared
to the field, because of the much higher environmental density (if the trends observed by \citealt{Lani13} can be extended to 
clusters). At the same time CRSGs also become less disk-dominated with decreasing redshift. The decreasing importance of
the disk component is  reflected in changes to all of the structural parameters, including the Sersic index $n$ and the axial ratios.
Similarly to \cite{Buitrago13} and \cite{Chang13} we observe an increasing contribution from disk-like components at $z >1$ 
and somewhat more oblate galaxies, with axial ratios closer to those of spiral disks (cf., \citealt{Rembold12,Cerulo14}).

The brighter and more massive galaxies may constitute a class apart. Brightest cluster galaxies, for instance, are often believed
to follow a more distinct evolutionary history than the less massive cluster population. We show the actual data in Fig. 3, 5 and 7 
for all objects. However, if we consider only the 10\% brighter (and likely more massive) systems, the values for $R_{eff}$ and $n$ 
are consistent with no significant evolution (i.e., these bright ellipticals appear to have reached their 'final' status early on in the 
history of the Universe) but the statistical power afforded by the small number of objects in the highest redshift samples does not 
suffice to draw strong conclusions on the evolution of these objects.

\subsection{A strawman model for morphological evolution}

Here we summarise a strawman model for this evolution which can explain the results described above, and  poses further 
useful questions on the details of this evolution. This model contains elements that will be familiar to anyone with a knowledge 
of the early studies of galaxy evolution \citep[e.g.][]{Larson1980}.  The model need not apply to each individual galaxy on the red
sequence, rather it describes the evolution necessary within a subset of CRSGs in order to produce the changes in the distributions 
of structural parameters over time.

At high redshift ($z>1$) we begin with a significant fraction of CRSGs having disk-like components that are both thin and have 
younger and more metal poor (bluer) stellar populations than their central spheroids. We associate the disk component with 
recently quenched star formation seen at $z>1.5$ \citep[e.g.][]{Tran2010}, so while it is bluer than the spheroid component, 
it is in the first stage of passive evolution and its photometric properties evolve rapidly. The model assumes a metallicity gradient 
(e.g., as for local samples of early-type galaxies -- \citealt{Rawle10,Kim13}) between the two morphological components, consistent 
with the many observations of radial metallicity gradients in disk galaxies \citep[e.g.][and references therein]{Moustakas2010}. In 
local disks, one observes metal abundance gradients of --0.2 to --0.6 dex per decade in radius (from the direct method of line ratios
in planetary nebulae and HII regions within resolved disks, as in \citealt{Moustakas2010}), which would yield colour gradients (from 
Fig.~\ref{models}) of the magnitude observed in local samples (such as the brighter members in Coma from \citealt{DenBrok11}). 
The effect of the metallicity difference on radial colour gradients is only evident at lower redshift as the difference in age of the bulge 
and disk components dominates at higher redshift.  

Given the constraint imposed on the overall colours of the galaxies by red sequence selection, the fraction of the mass provided
by any disk must be severely constrained, even when there is a colour gradient. Fig.~\ref{models}  shows the colour difference 
between two equal mass SSP bursts  with different formation times. Given the width of the rest-frame $B-R$ red sequence is 
only a few tenths of a magnitude and includes objects with a range of S\'ersic indices at all redshifts studied here, it is likely that 
any {\it young} disk component carries no more than $\sim 10$ per cent of the total stellar mass in a typical CSRG. Obviously 
any extended period of star formation in such a disk has less effect on the overall colour and therefore can contain more of the 
mass. \cite{Cen2014} shows how spiral galaxies evolve almost vertically in colour space and rapidly on to the red sequence 
once their star formation is terminated, in agreement with the scenario presented here, where such objects are present on the 
red sequence but still preserve their `spiral' morphologies, with thin disks.

With increasing time since quenching the disk fades and reddens faster than the older spheroidal population. After 3-4 Gyr the difference in
colours between the two components is largely dominated by the different metallicities and both fade and redden at approximately the 
same rate (as in Figure \ref{models}). This leads to the decreasing magnitude of and scatter in colour gradient with time. Over time the 
disk should also thicken to explain the changing axial ratio distribution for the population. This could simply be caused by the
quenched thin disk fading to reveal an underlying thick disk, or other internal and external processes could continually act to thicken the
disk. The precise mechanisms are beyond the scope of this paper and there may be several processes contributing to this, such as
disk instabilities or the addition of structure in the outskirts. However, simulations by \cite{Gnedin2003a,Gnedin2003b} 
demonstrated that the continual tidal heating suffered by disk galaxies subsequent to their infall into clusters was sufficient to thicken 
their disks by a factor of $\sim 2$ over their lifetime in the cluster, effectively transforming the shapes of late type spirals into those of 
S0s \citep{laurikainen2010}.  Over the same time, and for the same reason, the S\'ersic indices of those systems with the (initially) most
prominent disks change from $n<2$ to values more consistent with S0 galaxies, $2<n<3$, straightforwardly explaining the evolution 
seen in Fig.~\ref{Sersic}. A clear implication of this model is that we can link the high redshift CRSGs with relatively blue and thin 
disks to the S0 galaxies with small colour gradients and thicker disks seen to populate cluster red sequences at intermediate and low 
redshifts  at $\sim L^*$ and below. As the shapes of galaxies selected morphologically as E/S0 by \cite{Holden09} do not change
over time, our disk-like red sequence galaxies must naturally evolve to resemble present-day E/S0s. 

A key recent work on the low redshift cluster S0 population is that of \cite{head2014}, exploring the parameters of this population in
the Coma cluster. The simple model outlined here produces galaxies with photometric properties consistent with those determined 
by \cite{head2014}. Note that because S0 galaxies exist on the red sequences of clusters at all redshifts, the model does not imply 
that all low redshift cluster S0s formed in this manner between $0.6<z<1.4$, merely that the S0 population was significantly 
augmented over this time.

\begin{figure}
\includegraphics[width=0.5\textwidth]{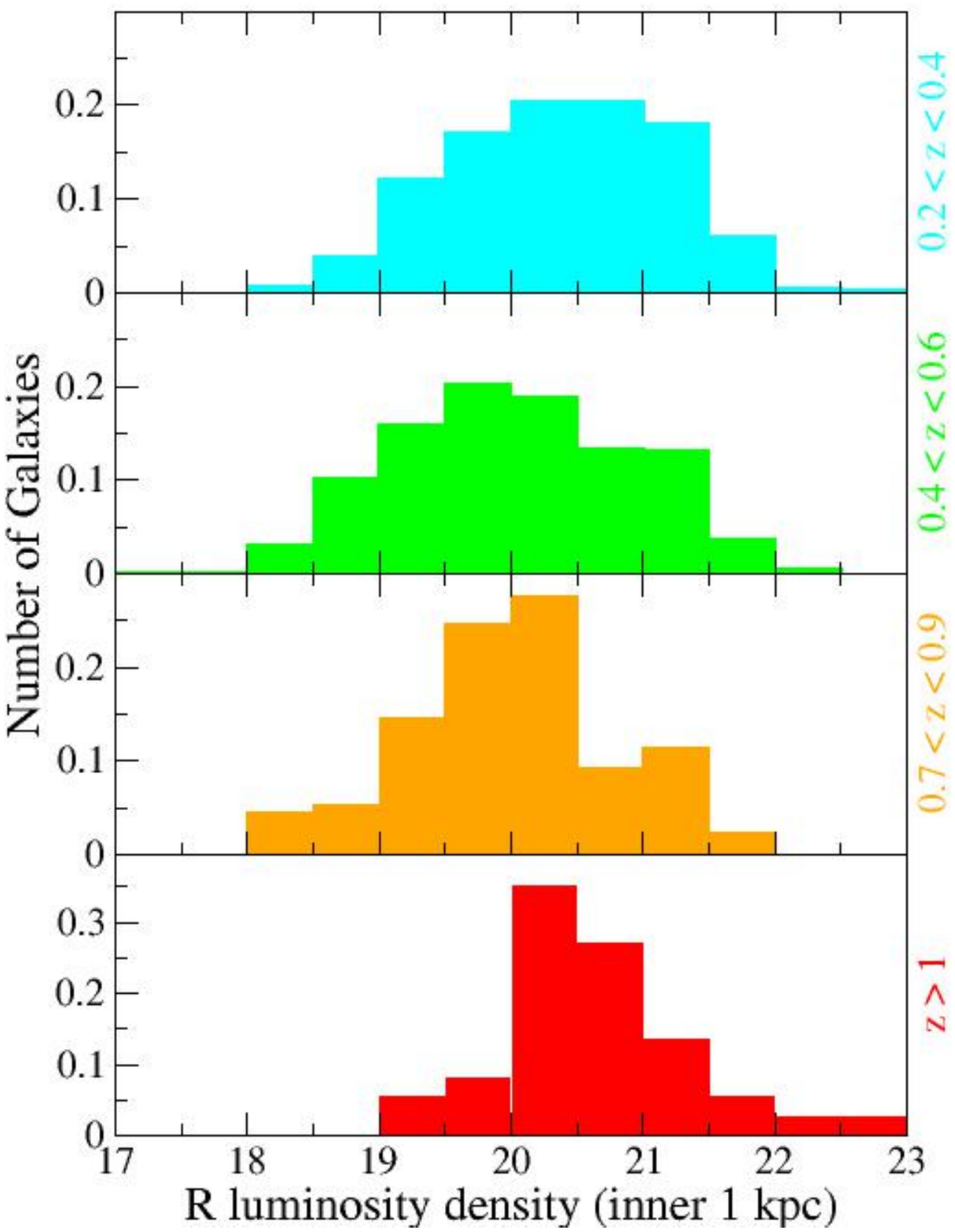}
\caption{Central luminosity density (within the inner kpc) for galaxies in our samples.
There is no strong evidence of an increase over $z < 1$ and therefore little evidence for bulge growth}
\label{concentration}
\end{figure}

\subsection{Limitations of the model and the current data}
There are several significant issues that are not constrained by this simple model and the existing data. The first is the reason for the
quenching of star formation in  disk components. This could simply be attributed to a loss of fuel after infall into the cluster. As ever,
suspects for such cluster-centric mechanisms include ram pressure stripping \citep[e.g.][]{Tonnesen2009,Cen2014}, virial shock heating
\citep{Birnboim2003} and strangulation \citep[e.g.][]{Larson1980, Peng2015} amongst others. Alternatively, it could be due to a secular
process, such as morphological quenching \citep{Martig2009,Huertas15}, where the spheroidal component of a galaxy grows to a size 
sufficient to stabilise the disk against further star formation. So long as star formation  in any disk is ended by $z\sim 1.5$, any of these
mechanisms may result in the systems at $1<z<1.4$ identified in Paper I having disk-like components.

A second issue is that of bulge growth. Although the model above was framed in terms of disk fading, because we are measuring the
evolution in relative strength of bulge and disk components with time there could equally be a r\^ole for bulge growth in this evolution. 
Work has recently been presented on the maturity of bulges and the quenching of star formation in $z\sim2$ star forming galaxies 
drawn from a range of environments \citep{Tacchella2015}. This indicated that these systems quench star formation from the inside 
out and that their central (bulge) surface densities are already comparable to those of  similar stellar mass $z=0$ early type galaxies 
with stellar masses above $10^{10}{\rm M}_\odot$, and within a factor of two of the $z=0$ values even for lower mass systems. The 
implication is that barring any subsequent mass growth through mergers, the bulges of galaxies with similar masses to those explored 
here are already largely in place by $z\sim 2$. The CRSGs explored in Paper I and here either originated in the field and
subsequently fell in to their clusters, or formed in an accelerated manner in existing high density regions as part of the original growth 
of the hosting clusters; it might be expected that their central spheroidal components are already in place before any disk component 
starts to quench. Given the already-noted limit to how much mass can be in any relatively young stellar disk given the narrow colour 
range in cluster red sequences, coupled with the lack of  growth of total stellar mass in the same galaxies since $z\sim 1.4$ (Paper I), 
it would seem that the bulges would have to be in place by $z\sim 1.4$. However, that does not account for any older disk component 
which could evolve morphologically, growing the bulge through inflow of stars and other material over time.

In order to test this we determine the luminosity in the $R$ band within the central 1 kpc ($\Sigma_1$) for each galaxy by integrating its fitted 
Sersic profile out to this point. This may be thought of as a proxy for the central density (as used by, e.g., \citealt{Fang2013} in their study of
low redshift transitioning `green valley' galaxies). Results are plotted as histograms in Fig~\ref{concentration}. The distribution of this `density;
measure is essentially the same for the `low' and `mid' redshift subsamples, implying no evolution in central density at $z<0.6$ (see also 
\citealt{vandokkum14}. The corresponding histogram for the `high' subsample shows if anything a higher typical central concentration. 
However, given only two clusters contribute, this could easily be attributable to shot noise from cluster to cluster. Only a small fraction of 
objects at this redshift have concentrations below the mean, suggesting that the inner regions of the galaxy population have been built 
by this epoch. The `very high' sample may be different from the others: there is a relative lack of bright `bulges' but the small number
statistics and the increased uncertainty in each measurement caused by the relatively large size of the WFC3/$H_{F160W}$ PSF 
in comparison to the effective radii of galaxies make interpretation of this plot difficult without more data. 

Consequently, it appears that the high central densities of the spheroidal components of the CRSGs are in place by $z\sim 1$. 
The origin of the outer, younger, more metal poor material is more uncertain: as an example, it may itself have been accreted through 
minor mergers at an earlier epoch (before $z=1.5$), perhaps even during pre-processing of galaxies before they join the main cluster.
From both the limited results on concentration obtained here and the arguments based on the work of \cite{Tacchella2015}, it is likely 
that the spheroidal components of most of the $1<z<1.4$ CRSGs are already in place. This would imply that the decrease in the 
contribution from disk-like components in the population over the 3-4 Gyr between $z\sim 1.4$ and $z\sim 0.5$ was mainly or solely 
due to the fading or rearrangement of the disks and not from the buildup of a bulge component (or at least the central regions of 
any bulge). This evolution in the disks may be triggered by disk instabilities, tides, harassment, fly-bys and a series of other 
external and internal processes we are not in a position to explore here. A further indication that the morphological transformation
may be due to changes in the disk, is the concurrent evolution observed in the axial ratios and S{\'e}rsic indices, which
tend to reinforce our impression that the spheroidal portion of these galaxies has been established at early times and only
the younger, and perhaps more fragile, disks appear to be involved in the morphological changes we observe.

One result that does not directly follow from the strawman model is the (albeit small)  increase in the typical size ($R_{eff}$) of 
the CRSG population  between $z\sim1.4$ and $z\sim0.6$. If a disks fades and reddens relative to its bulge, with no large-scale 
change to the bulge structure it would be expected that the overall $R_{eff}$ would probably decrease (albeit by a modest amount)
rather than increase, in the absence of any other morphological evolution. Given the small size of any evolution, it could be explained 
by other evolutionary effects in only a subset of the comparatively small `high' and `very high' subsamples. We note here that 
the strawman model does not account for this potential behaviour, but is certainly not currently ruled out by it.  For instance,
it is possible that any size evolution in the overall CRSG population could be driven by galaxies other than those which are 
evolving morphologically in the manner discussed here.

Alternatively, one may envisage the disk being disturbed, for instance by minor mergers or fly-bys, and then added
on to the outer regions of the bulge, thus driving a modest increase in size and S{\'e}rsic index as structure is increased in the outer
regions, and reducing the oblateness of galaxies. Such fly-bys would be frequent in the dense environment of clusters and may result
in significant structural changes in halos \citep{Sinha15}. Disk instability would be another possibility (e.g., \citealt{Shankar13,
Huertas15}). In the model by \cite{Naab09}, for instance, galaxies are born as highly flattened objects and their disks are later 
on destroyed or rearranged by interactions and mergers with minor satellites (e.g., \citealt{Quilis12}) to form a more spheroid-dominated
population. Several other mechanisms are possible, as discussed in the vast literature on size growth of spheroids, but we are not
in a position here to speculate.

A clear improvement to the analysis presented here would be to carry out detailed two-component surface brightness fitting to CRSGs 
over the entire redshift range and perhaps beyond $z\sim 1.4$ where the analysis may need to be extended to slightly bluer galaxies 
due to the increased influence of star formation in red sequence progenitors. This is not reliably achievable with the current HST data,
especially at high redshift where the most changes are seen (cf. \citealt{Cameron2007}). Deeper exposures could be obtained with HST. 
However,  an equally important issue is to be able to resolve the galaxies on scales smaller than $R_{eff}$ in the reddest bands. Higher spatial 
resolution will allow the analysis to separate out any disk and bulge component and therefore to securely determine how much of the observed 
evolution is caused by bulge growth and how much to disk fading. This will be straightforwardly achievable with JWST.

 \section{Conclusions}
 
We have explored the morphological evolution of cluster red sequence galaxies using a large sample of such galaxies drawn from 
multiple clusters over the redshift range $0.2<z<1.4$ split into four subsamples by redshift. Our previous work showed there were 
significant differences between the populations at $1<z<1.4$ and  $z \sim 0$. Here we fill in details of their evolution over
the intervening 9-10 Gyr. Our major findings are:
 
\begin{itemize}

\item We confirm that the distribution in structural parameters of high redshift cluster galaxies differs from those at lower redshift. 
At least a proportion of the population evolves morphologically having arrived on the red sequence subsequent to quenching of 
star formation.
 
 \item The measurements indicate that this evolution  largely ends by $z\sim 0.5-0.6$, a lookback time of 5-6 Gyr and a likely 3-4 Gyr 
after quenching of star formation. This evolution is seen in the distribution of S\'ersic parameters, axial ratios, sizes and colour gradients. 
 
 \item  The evolution in each of these parameters indicates typical objects become less disk-dominated with time, with the strength of any colour
gradient (bluer on the outside)  decreasing similarly. The data broadly fit a simple model where (thin) disk components fade and redden 
more quickly than their central bulges, straightforwardly explained by more recent quenching of star formation in the disk than in the
bulge. The timescale for this matches the $3-4$ Gyr  over which we can detect changes in the structural parameters. During this time
the disks are required to thicken in order to explain the changing axial ratio distribution, agreeing with the timescale for the expected
tidal thickening of disks in clusters. The obvious implication of the model is that these initially disk-dominated galaxies are the progenitors of the
passive S0 systems that populate (indeed dominate) the red sequence of intermediate and low redshift clusters at luminosities 
around $L^*$ and below (also see \citealt{Chan2016,Lopes16}).
 
 \item We confirm that the size growth in the population of red sequence galaxies in clusters is significantly less over the past $9-10$
Gyr than that of similar mass passively-evolving field galaxies. The data indicate that any growth is likely to occur at $z>0.6$ as the
size distributions of our two lower redshift subsamples appear indistinguishable.
 
 \item While the decreasing prominence of any disk component could have a contribution from bulge growth over time, consideration
of the fraction of the luminosity (mass) within the central 1kpc of each galaxy indicates that at least to $z\sim 1$ this is not a 
significant factor in the evolution of the measured structural parameters.
 
 \item Although further similar HST data sets  (with both rest-frame $B-$ and $R-$band imaging) for $z>0.6$ clusters would improve 
the statistical significance of some of these results, improvements in the interpretation of these results will only come from the ability 
to securely fit multi-component surface brightness profiles  to high redshift samples. This will only be achievable through JWST imaging
which will give sufficient signal-to-noise and spatial resolution at the highest redshifts to separate the co-evolution of disk and bulge components 
in cluster red sequence galaxies.
 
 \end{itemize}

\section*{Acknowledgments}

This work is based on observations made with the NASA/ESA Hubble Space Telescope, and obtained from the Hubble Legacy Archive, 
which is a collaboration between the Space Telescope Science Institute (STScI/NASA), the Space Telescope European Coordinating
Facility (ST-ECF/ESA) and the Canadian Astronomy Data Centre (CADC/NRC/CSA). Some of the data presented in this paper were 
obtained from the Mikulski Archive for Space Telescopes (MAST). STScI is operated by the Association of Universities for Research in Astronomy, Inc., 
under NASA contract NAS5-26555.

\label{lastpage}

\bsp

\end{document}